\documentclass [aps,prb,preprint,floatfix,showpacs,superscriptaddress]{revtex4}

\usepackage {amsmath,amssymb,epsfig,color,tabularx}

\begin{document}

\title
{The semiconductor-to-ferromagnetic-metal transition in FeSb$_2$}

\author{A.V.~Lukoyanov}
\affiliation{Institute of Metal Physics, Russian Academy of
Sciences-Ural Division, 620041 Yekaterinburg GSP-170, Russia}
\affiliation{Ural State Technical University-UPI,
620002 Yekaterinburg, Russia}
\author{V.V.~Mazurenko}
\affiliation{Institute of Metal Physics, Russian Academy of
Sciences-Ural Division, 620041 Yekaterinburg GSP-170, Russia}
\affiliation{Ural State Technical University-UPI,
620002 Yekaterinburg, Russia}
\author{V.I.~Anisimov}
\affiliation{Institute of Metal Physics, Russian Academy of
Sciences-Ural Division, 620041 Yekaterinburg GSP-170, Russia}
\author{M.~Sigrist}
\affiliation{Institut f\"ur Theoretische Physik, ETH-H\"onggerberg, 
CH-8050 Z\"urich, Switzerland}
\author{T.M.~Rice}
\affiliation{Institut f\"ur Theoretische Physik, ETH-H\"onggerberg, 
CH-8050 Z\"urich, Switzerland}

\date{\today}

\begin{abstract}
We propose FeSb$_2$ to be a nearly ferromagnetic small gap semiconductor, hence a direct analog 
of FeSi. We find that despite different compositions and crystal structures, 
in the local density approximation with on-site Coulomb repulsion correction (LDA+$U$) method 
magnetic and semiconducting solutions for $U$=2.6~eV are energetically degenerate 
similar to the case of FeSi. For both FeSb$_2$ and FeSi (FeSi$_{1-x}$Ge$_x$ alloys) 
the underlying transition mechanism allows one to switch from a small gap semiconductor 
to a ferromagnetic metal with magnetic moment $\approx 1 \mu_B$ per Fe ion with external magnetic field. 
\end{abstract}

\maketitle
The unusual crossover from a small gap semiconductor at low temperatures 
to a metallic state with enhanced magnetic fluctuations above room temperature 
observed in FeSi has long been a subject 
of great interest~\cite{FeSb2_1,Kondo,Anisimov96,semicond2,Jarlborg04}. 
Two different models have been proposed to explain this anomaly. One proposal 
is that FeSi is a Kondo insulator~\cite{Kondo}. The second is that FeSi is a nearly 
ferromagnetic semiconductor~\cite{Anisimov96,semicond2}. This latter proposal is supported by 
{\it ab initio} electronic structure calculations using the LDA+$U$ method~\cite{LDA+U} 
which found that a ferromagnetic metallic state was very close by 
in energy~\cite{Anisimov96}. Further support for this second model comes from 
the direct observation of this semiconductor-metal transition as the lattice 
is expanded by the isoelectronic substitution of Ge for Si~\cite{Anisimov02}. 
In order to determine critical magnetic field a minimal two-band model with interband 
interaction was suggested. Even in the mean field approximation the model nicely describes 
the full phase diagram of the FeSi$_x$Ge$_{1-x}$ alloy series~\cite{Anisimov02}.

Recently, a second Fe-compound, FeSb$_2$, was found to have a similar crossover 
as FeSi from a small gap semiconductor to a metallic state with strong magnetic 
fluctuations~\cite{FeSb2,Petrovic03}. This raises immediately the question whether 
{\it ab initio} calculations confirm the close analogy between the two Fe compounds. 
This is a nontrivial question since these electronic structure calculations show 
strong hybridization between the Fe-3$d$ orbitals and the $s$-$p$ electrons 
of the close Si or Sb neighbors. This strong hybridization clashes with 
the assumption of weak hybridization between localized and itinerant electrons 
that underlies a Kondo insulator description and makes it difficult to relate 
the Kondo insulator model to the {\it ab initio} electronic structure. 
However in view of the differing compositions and crystal structures 
of FeSi and FeSb$_2$ it is by no means 
obvious that closely similar models can be derived from {\it ab initio} 
electronic structure calculations for both compounds. For this reason it is 
clearly important to examine FeSb$_2$ closely.

FeSb$_2$ crystallizes in the marcasite crystal structure~\cite{structure}. Each Fe atom 
is surrounded by a slightly distorted octahedron of Sb neighbors with 
2 neighbors at 2.57 \AA~and 4 at 2.59 \AA. The octahedra are corner sharing in 
the $ab$ plane and edge sharing along the $c$-axis. 

\begin{figure}
\begin{center}
\includegraphics[scale=.35]{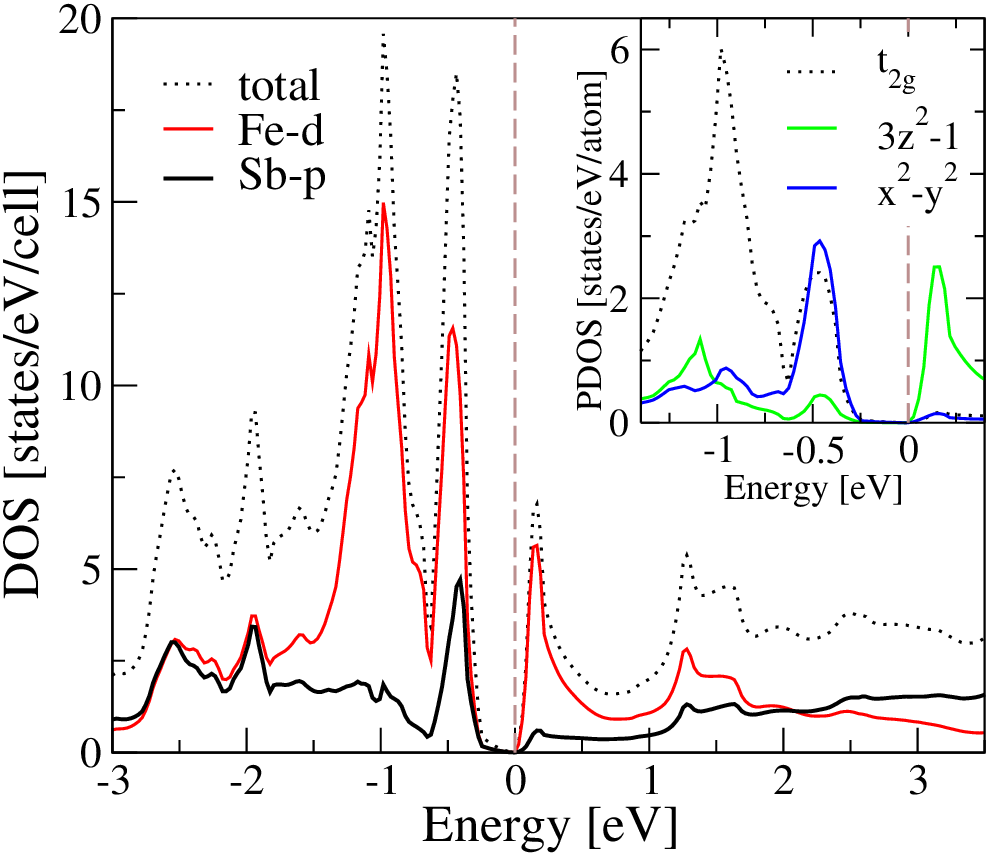}
\caption{Total and partial densities of states for FeSb$_2$ 
from the LDA calculation. Inset shows partial $t_{2g}$-DOS 
and $3z^2-r^2$, $x^2-y^2$ orbitals DOS of Fe-3$d$ states. 
The Fermi energy corresponds to zero.}
\label{fig:DOSfesb2}
\end{center}
\end{figure}

\begin{figure}
\begin{center}
\includegraphics[scale=.35]{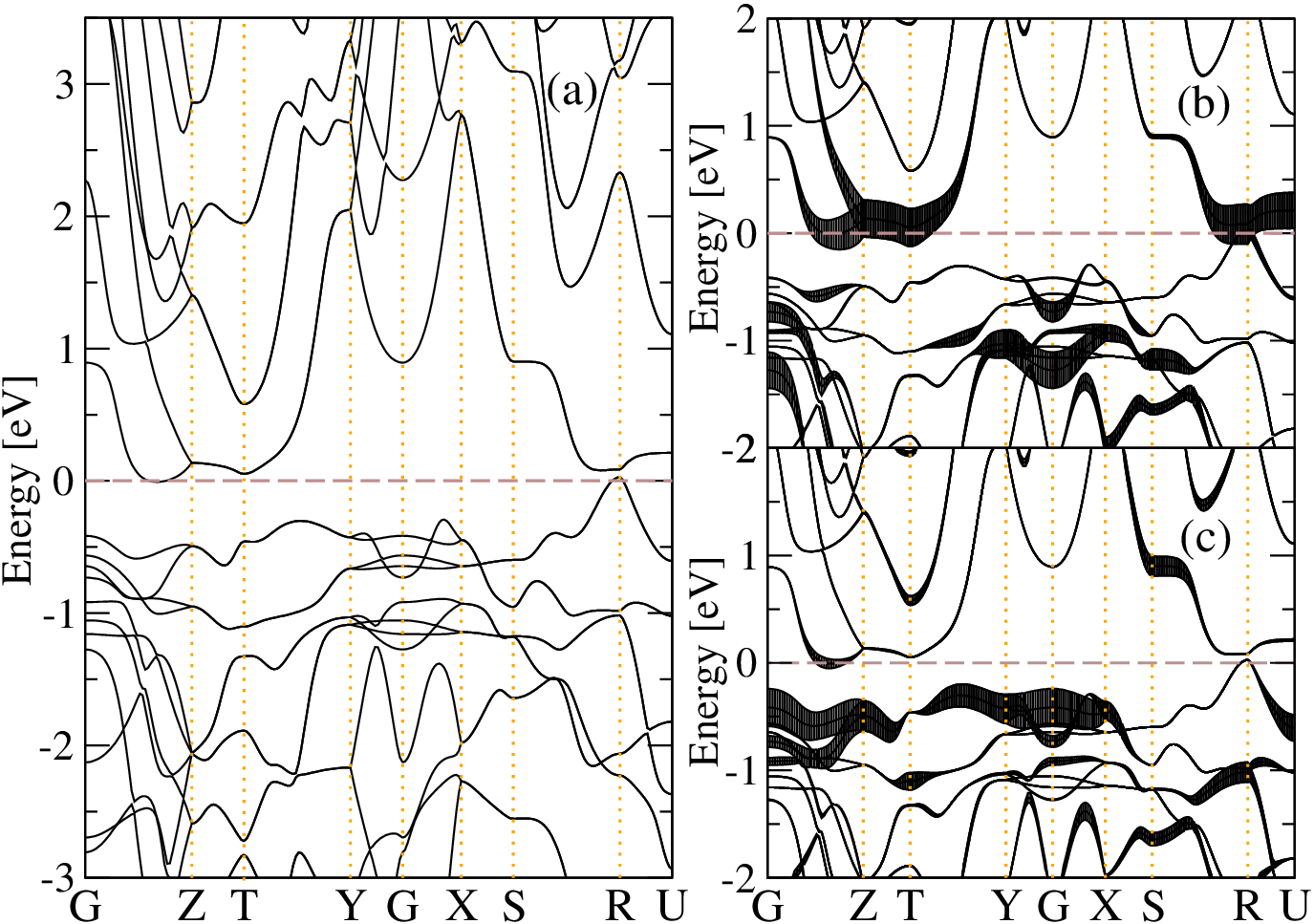}
\caption{(a) Band structure of FeSb$_2$ from 
the LDA calculation. Right panels show partial 
contributions of (b) $3z^2-r^2$ and (c) $x^2-y^2$ orbitals 
to the total band structure. Additional broadening of the bands corresponds 
to the contribution of the orbital. The Fermi energy corresponds to zero.}
\label{fig:BANDSfesb2}
\end{center}
\end{figure}

The results of density functional calculations using a local density approximation 
(LDA)~\cite{DFT} within TB-LMTO-ASA program~\cite{LMTO} are shown in Figures~\ref{fig:DOSfesb2} 
and \ref{fig:BANDSfesb2} (the calculations were performed without spin-polarization). 
Atomic spheres radii were chosen as R(Fe) = 2.67 a.u. and R(Sb) = 2.99 a.u., in ASA
approximation to fulfil the volume 6 classes of empty spheres were inserted
with radii from 1.56 to 0.81 a.u.
The overall bandwidth is $\approx$10~eV with strong Fe-Sb hybridization due to the short Fe-Sb bonds. 
The density of states (DOS) is shown in Figure~\ref{fig:DOSfesb2}. The Fermi energy 
lies in a small band gap similar to the case of FeSi. Also in Figure~\ref{fig:DOSfesb2} 
the DOS broken down in Fe-3$d$($t_{2g}$) and Sb-4$p$ states is displayed. 
Narrow bands appear above and below the band gap with predominantly 3$d$ character -- 
again similar to the case of FeSi. There is however a substantial contribution of 
the Sb-4$p$ states to valence band peak just below the Fermi energy but in the 
interval from the Fermi energy to $\approx$1~eV Sb-4$p$ states are presented very weakly.

The band structure of FeSb$_2$ is presented in Figure~\ref{fig:BANDSfesb2}(a). There is 
a small indirect energy gap with the minimum in the conduction bands lying on the 
GZ-lines and the maxima of the valence band is at the R-point. 
The narrow peak in the DOS at the bottom of the conduction band arises from the flat 
bands that extend over roughly half of the Brillouin near the Z- and R-points. 
The orbital character of these flat bands is predominantly $3z^2-r^2$ (in the global 
coordinate system) as illustrated in Figure~\ref{fig:BANDSfesb2}(b). 
These are 3$d$ orbitals pointing away from the Sb-sites in the octahedra 
surrounding the Fe atoms. The top of the valence band lies mostly in the DOS peak 
$\approx$0.3~eV below the conduction band and has predominantly 3$d$ 
($x^2-y^2$)-character. There is however a small density of states with mostly Sb-$p$ 
character which is responsible for the small indirect band gap. 

The main feature of the LDA band structure is this small gap ($\approx$0.3~eV) 
between relatively flat bands of 3$d$ character giving rise to sharp 
peaks in the DOS. But the Sb-4$p$ band contributes small density of states 
in this gap with zero at the Fermi level. 
In reference~\cite{opticalgap} from measurements of optical 
reflectivity the semiconducting optical gap was estimated as E$_g$=0.035 eV. 
This value is larger than that obtained from the resistivity measurements 
$\approx$0.02-0.025 eV (see reference~\cite{Petrovic03}). This situation is similar 
to FeSi except that E$_g$ does not show any dependence on temperature up to 
the crossover temperature. 

The LDA approximation was extended to allow for spin and orbital 
polarization by Anisimov {\it et~al.}~\cite{LDA+U} by introducing a local Coulomb 
repulsion, $U$. This LDA+$U$ method when applied to FeSi found that there was a ferromagnetic 
metallic state very close by in energy to the small gap semiconducting state of the LDA. 
The exact energy difference between the two depends on the choice of $U$. 
This result of LDA+$U$ is nicely confirmed by the fact that the isostructural-isoelectronic 
compound FeGe has the ferromagnetic metallic ground state. Actually if spin-orbit terms 
were included the ferromagnetism of FeGe would develop a long period spiral structure 
as observed in practice due to the presence of a Dzyaloshinskii-Moriya term in resulting 
from the absence of inversion about the Fe sites. 

The phase diagram of the FeSi$_x$Ge$_{1-x}$ alloys can be reproduced by choosing 
the reasonable value of $U$, an onsite Coulomb repulsion on the Fe sites 
of $U$=3.7~eV~\cite{Anisimov02}. 

\begin{figure}
\begin{center}
\includegraphics[scale=.35]{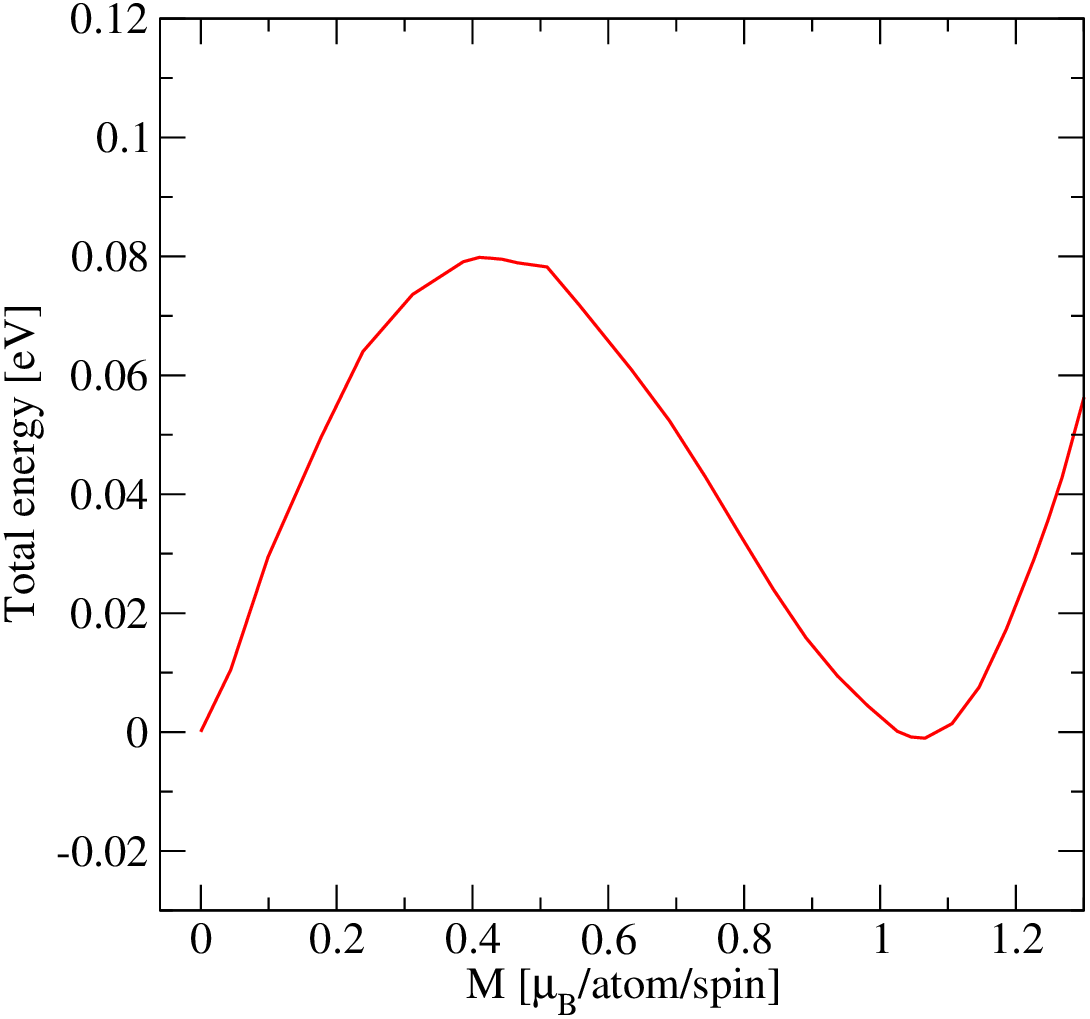}
\caption{Total energy dependence on magnetization 
value for FeSb$_2$. From the LDA+$U$ calculation 
with $U$=2.6~eV and $J_H$=0.88~eV. Energy is related 
to the energy of nonmagnetic state.}
\label{fig:MAGNfesb2}
\end{center}
\end{figure}

We have applied the LDA+$U$ method to FeSb$_2$. As in the case of FeSi a second local 
minimum appears in the energy vs. uniform magnetization at a value of 1 $\mu_B$ per Fe 
(see Figure~\ref{fig:MAGNfesb2}). In this set of calculations we performed fixed spin moment
procedure~\cite{Anisimov96}. Again the exact energy difference between the two 
minima is dependent on the value of $U$. 

In all our calculation we used the value $J_H$=0.88 eV for the exchange (Hund's) Coulomb 
parameter but varied the value of direct Coulomb parameter $U$. As a result, in FeSb$_2$ 
we found the critical value of the direct Coulomb parameter $U_c$=2.6 eV. 
As in FeSi and FeSi$_x$Ge$_{1-x}$ alloys, for any value of the $U$ parameter less than $U_c$ 
it is the nonmagnetic ground state that is lower in total energy. Otherwise for the $U$ values 
above $U_c$ ferromagnetic state is lower in energy. Only for $U_c$ these two states have the same energy.

\begin{figure}
\begin{center}
\includegraphics[scale=.35]{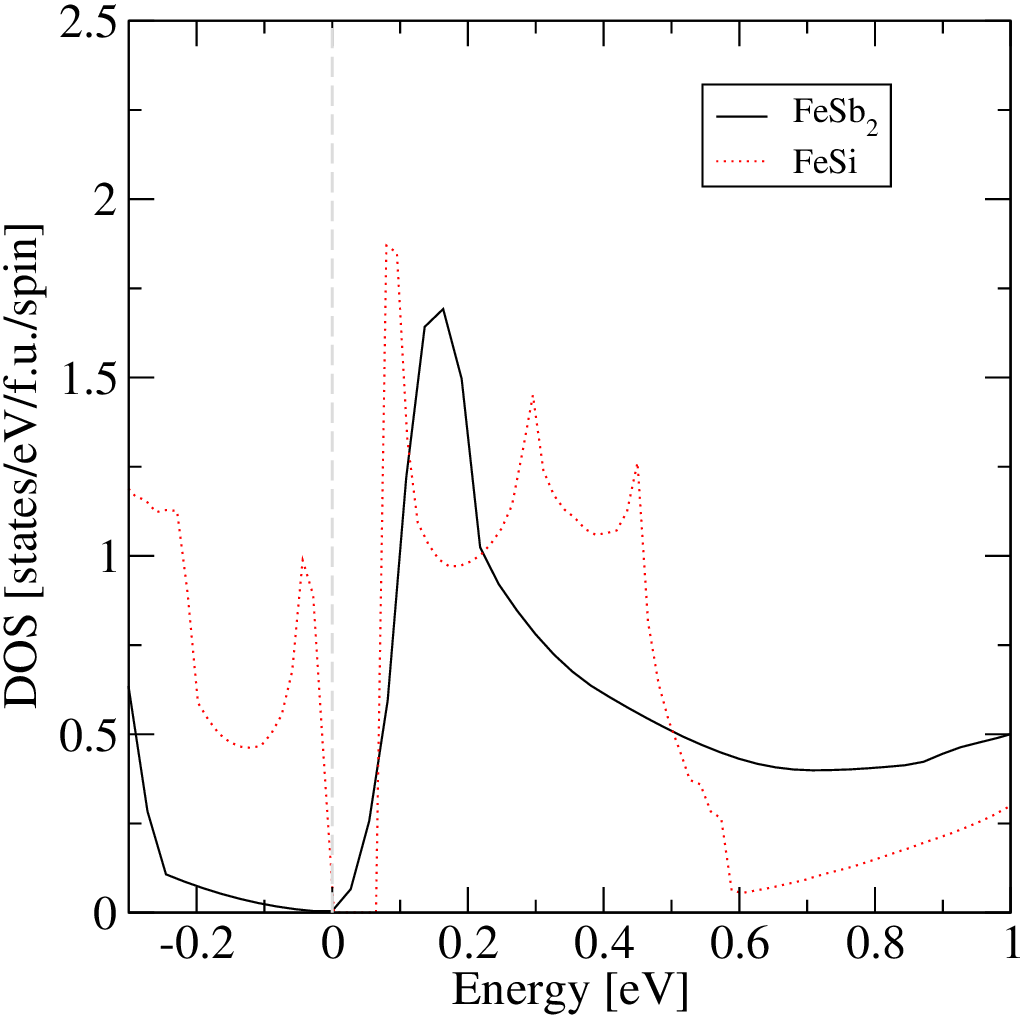}
\caption{Total densities of states of FeSb$_2$ 
 and FeSi~\cite{Anisimov96} from the LDA 
 calculations near the Fermi energy.}
\label{fig:DOScomp}
\end{center}
\end{figure}

Comparison of the total FeSb$_2$ density of states near the Fermi energy to FeSi 
is shown in Figure~\ref{fig:DOScomp}. 

The close similarity between the LDA+$U$ results for FeSi and FeSb$_2$ agrees well with 
the close correspondence in their properties. It demonstrates the nearly ferromagnetic 
character of these small gap semiconductors.

\section{Acknowledgments}
The authors are grateful to C.~Petrovic for stimulating conversations. Three of the authors 
(A.V.L., V.V.M. and V.I.A.) are grateful to the Institut f\"ur Theoretische Physik, ETH Z\"urich, 
for the hospitality and the last author (T.M.R.) thanks Brookhaven National Laboratory for their hospitality. 
This work was supported by Russian Foundation for Basic Research under the grant RFFI-04-02-16096, 
by Netherlands Organization for Scientific Research through NWO 047.016.005, the Center for 
Theoretical Studies of ETH Z\"urich and the NCCR of MaNEP.
A.V.L. acknowledges support from the Dynasty Foundation and International Center for Fundamental Physics in Moscow. 
V.V.M. thanks INTAS Program Ref. No. 04-83-3230 and grant program of President of Russian Federation, 
Grant No. MK-15-73.2005.2. Electronic structure calculations have been performed on the Gonzales cluster, ETH Z\"urich.


\newpage

\end{document}